\documentclass[%
reprint,
superscriptaddress,
 amsmath,amssymb,
prc,
]{revtex4-1}

\usepackage[dvipdfmx]{graphicx}
\usepackage{tablefootnote}
\usepackage{graphicx}% Include figure files
\usepackage{dcolumn}% Align table columns on decimal point
\usepackage{bm}% bold math
\usepackage{here}

\usepackage{color}

\begin{document}

%\preprint{APS/123-QED}

\title{Alpha-decay-correlated mass measurement of \textcolor{black}{$^{206,207g,m}$}Ra using an $\alpha$-TOF detector equipped \textcolor{black}{multi-reflection time-of-flight mass spectrograph} system}% isotope}% Force line breaks with \\
%\thanks{A footßnote to the article title}%

\author{T.~Niwase}
\email{tniwase@post.kek.jp}
\affiliation{KEK Wako Nuclear Science Center, Wako, Saitama 351-0198, Japan}
\affiliation{RIKEN Nishina Center for Accelerator-Based Science, Wako, Saitama 351-0198, Japan}
\affiliation{Department of Physics, Kyushu University, Nishi-ku, Fukuoka 819-0395, Japan}

\author{M.~Wada}
\affiliation{KEK Wako Nuclear Science Center, Wako, Saitama 351-0198, Japan}

\author{P~Schury}
\affiliation{KEK Wako Nuclear Science Center, Wako, Saitama 351-0198, Japan}

\author{P.~Brionnet}
\affiliation{RIKEN Nishina Center for Accelerator-Based Science, Wako, Saitama 351-0198, Japan}

\author{S.~D.~Chen}
\affiliation{%
Department of Physics, The University of Hong Kong, Hong Kong
}%
\affiliation{KEK Wako Nuclear Science Center, Wako, Saitama 351-0198, Japan}

\author{T.~Hashimoto}
\affiliation{%
Institute for Basic Science, 70, Yuseong-daero 1689-gil, Yusung-gu, Daejeon, Korea
}%

\author{H.~Haba}
\affiliation{RIKEN Nishina Center for Accelerator-Based Science, Wako, Saitama 351-0198, Japan}

\author{Y.~Hirayama}
\affiliation{KEK Wako Nuclear Science Center, Wako, Saitama 351-0198, Japan}

\author{D.~S.~Hou}
\affiliation{Institute of Modern Physics, Chinese Academy of Science, Lanzhou 730000, China}
\affiliation{University of Chinese Academy of Science, Beijing 100049, China}
\affiliation{School of Nuclear Science and Technology, Lanzhou University, Lanzhou, 730000, China}

\author{S.~Iimura}
\affiliation{Department of Physics, Osaka University, Osaka, Japan}
\affiliation{RIKEN Nishina Center for Accelerator-Based Science, Wako, Saitama 351-0198, Japan}
\affiliation{KEK Wako Nuclear Science Center, Wako, Saitama 351-0198, Japan}

\author{H.~Ishiyama}
\affiliation{RIKEN Nishina Center for Accelerator-Based Science, Wako, Saitama 351-0198, Japan}

\author{S.~Ishizawa}
\affiliation{Graduate School of Science and Engineering, Yamagata University, Yamagata, Japan}
\affiliation{RIKEN Nishina Center for Accelerator-Based Science, Wako, Saitama 351-0198, Japan}

\author{Y.~Ito}
\affiliation{%
Japan Atomic Energy Agency(JAEA), Tokai, Ibaraki, Japan
}%

\author{D.~Kaji}
\affiliation{Department of Physics, Kyushu University, Nishi-ku, Fukuoka 819-0395, Japan}

\author{S.~Kimura}
\affiliation{RIKEN Nishina Center for Accelerator-Based Science, Wako, Saitama 351-0198, Japan}

\author{J.~Liu}
\affiliation{%
Department of Physics, The University of Hong Kong, Hong Kong
}%
\affiliation{KEK Wako Nuclear Science Center, Wako, Saitama 351-0198, Japan}

\author{H.~Miyatake}
\affiliation{KEK Wako Nuclear Science Center, Wako, Saitama 351-0198, Japan}

\author{J.~Y.~Moon}
\affiliation{%
Institute for Basic Science, 70, Yuseong-daero 1689-gil, Yusung-gu, Daejeon, Korea
}%

\author{K.~Morimoto}
\affiliation{RIKEN Nishina Center for Accelerator-Based Science, Wako, Saitama 351-0198, Japan}

\author{K.~Morita}
\affiliation{Department of Physics, Kyushu University, Nishi-ku, Fukuoka 819-0395, Japan}
\affiliation{Research Center for SuperHeavy Elements, Kyushu University, Nishi-ku, Fukuoka 819-0395, Japan}

\author{D.~Nagae}
\affiliation{Department of Physics, Kyushu University, Nishi-ku, Fukuoka 819-0395, Japan}
\affiliation{Research Center for SuperHeavy Elements, Kyushu University, Nishi-ku, Fukuoka 819-0395, Japan}

\author{M.~Rosenbusch}
\affiliation{KEK Wako Nuclear Science Center, Wako, Saitama 351-0198, Japan}

\author{A.~Takamine}
\affiliation{Department of Physics, Kyushu University, Nishi-ku, Fukuoka 819-0395, Japan}

\author{T.~Tanaka}
\affiliation{Department of Nuclear Physics, Research School of Physics, The Australian National University, Canberra, Australian Capital Territory 2601, Australia}

\author{Y.~X.~Watanabe}
\affiliation{KEK Wako Nuclear Science Center, Wako, Saitama 351-0198, Japan}

\author{H.~Wollnik}
\affiliation{New Mexico State University, Las Cruces, NM 88001, USA}

\author{W.~Xian}
\affiliation{%
Department of Physics, The University of Hong Kong, Hong Kong
}%
\affiliation{KEK Wako Nuclear Science Center, Wako, Saitama 351-0198, Japan}

\author{S.~X.~Yan}
\affiliation{Institute of Mass Spectrometer and Atmospheric Environment, Jinan University,Guangzhou, 510632, China}


\date{\today}% It is always \today, today,
\begin{abstract}

The atomic masses of the isotopes $^{206,207}$Ra have been measured via decay-correlated mass spectroscopy using a multi-reflection time-of-flight mass spectrograph equipped with an $\alpha$-TOF detector. The Ra isotopes were produced as fusion-evaporation products in the $^{51}$V+$^{159}$Tb reaction system and delivered by the gas-filled recoil ion separator GARIS-II at RIKEN. The $\alpha$-TOF detector provides for high-accuracy mass measurements by correlating time-of-flight signals with subsequent $\alpha$-decay events. \textcolor{black}{The masses of $^{206}$Ra and $^{207g,m}$Ra} were directly measured using a multi-reflection time-of-flight mass spectrograph equipped with an $\alpha$-TOF detector.  \textcolor{black}{The mass excesses of $^{206,207g}$Ra and the excitation energy of $^{207m}$Ra were determined to be $M\!E$=3540(50)~keV/c$^2$, 3538(15)~keV/c$^2$, and $E_{\rm ex}$~=~552(42)~keV,  respectively. }The $\alpha$-decay branching ratio of $^{207m}$Ra, $b_{\alpha}=0.26(20)$, was directly determined from decay-correlated time-of-flight signals, and the reduced alpha width of $^{207m}$Ra was calculated to be $\delta^{2}=43^{+68}_{-34}$~keV from the branching ratio. The spin-parity of $^{207m}$Ra was confirmed to be $J^{\pi}$ = 13/2$^-$ from decay correlated mass measurement results.

%\begin{description}
%\end{description}
\end{abstract}

\pacs{Valid PACS appear here}% PACS, the Physics and Astronomy
\maketitle

%\tableofcontents

%Introduction

\section{Introduction}
The structure of heavy and superheavy nuclides are strongly influenced by shell effects, and investigations of the ground and excited state properties provide us important information for the understanding of these nuclides. The binding energy of the ground state nuclide, as a direct mapping of the shell structure, is best determined by atomic mass measurements. However, for many nuclides the atomic masses have only been indirectly determined through decay and reaction $Q$-values. To preclude possible errors, which can compound in long decay chains, requires direct measurements. An exemplary case of such was the direct mass measurement of $^{150}$Ho by Penning-trap mass spectrometry at CERN/ISOLDE, wherein they found that the indirect measurements, from by $\beta$-decay spectroscopy, had an 800~keV discrepancy due to a misidentification of the excited and ground states~\cite{Lunney_150Ho,Beck_150Ho}. 
 \par	The structure of $\gamma$-transitioning isomers has been studied with high precision based on $\gamma$-ray spectroscopy.  However, there are still many nuclides in which the isomeric states undergo direct $\alpha$- or $\beta$-decay, precluding the application of $\gamma$-ray spectroscopy.  Separately, there have been a number of decay spectroscopy measurements assisted by mass separation, utilizing both Penning traps and multi-reflection time-of-flight mass spectrographs (MRTOF-MS).  The first experiment where such trap-assisted decay spectroscopy was performed with REXTRAP at CERN/ISOLDE for conversion electron studies~\cite{Weissman}. Since then, several experiments have been performed using \textcolor{black}{high-resolution} mass separators coupled with a decay station~\cite{Kowalska,Lorenz,Timo,Althubiti,RintaAntila}. Apart from an effort to measure half-lives using variable storage time in an ion trap connected to an MRTOF~\cite{Wolf}, until now when performing trap-assisted decay spectroscopy the mass determinations and decay measurements were performed independently; the mass spectrometers have always been employed as high-resolution mass separators. 
 \par Recently, we have developed a novel detector, which we refer to as an $\alpha$-TOF~\cite{Niwase}, that can simultaneously measure ion implantation (to deduce time-of-flight) and subsequent $\alpha$-decay events from implanted ions.  From these correlated signals, we can perform mass and decay spectroscopy for multiple nuclides simultaneously.  We have previously described the use of this decay-correlated mass spectroscopy to suppress background.  Other nuclear properties can be inferred using the detector; the life-time of each nuclide, for instance, can be determined from the time intervals between the time-of-flight signal and the decay signal. In this paper, we report on the decay-correlated mass spectroscopy of $^{206,207}$Ra using an MRTOF-MS equipped with an $\alpha$-TOF detector, demonstrating several capabilities beyond atomic mass determination.
% , allowing for correlated  with the mass of the nuclides, and have discussed the possibility of simultaneous mass spectroscopy and decay spectroscopy measurements by offline measurements and numerical simulations. This technique allows us to detect the decay signal each correlated with the time-of-flight (TOF) signal as long as the life-time is sufficiently shorter than the counting rate. From these correlated signals, we can perform mass and decay spectroscopy for multiple nuclides simultaneously. The life-time of each nuclide can also be determined from the time intervals between the time-of-flight signal and the decay signal.

	%In this paper, we report on the decay correlated mass spectroscopy of $^{206,207}$Ra isotopes
	%have measured $^{206,207}$Ra as a demonstration of decay correlated mass spectroscopy 
	%using a MRTOF-MS equipped with an $\alpha$-TOF detector. % at the RIKEN Gas-filled Recoild Ion Separator (GARIS-II).
	%In the following section, the experimental setup and data analysis method will be described and the results are compared with the atomic mass evaluation (AME2016) and the nuclear data sheets.

\section{Experiments}
\begin{figure}[t]
%	\vspace{2mm}
	\centering
	\includegraphics[scale=0.57]{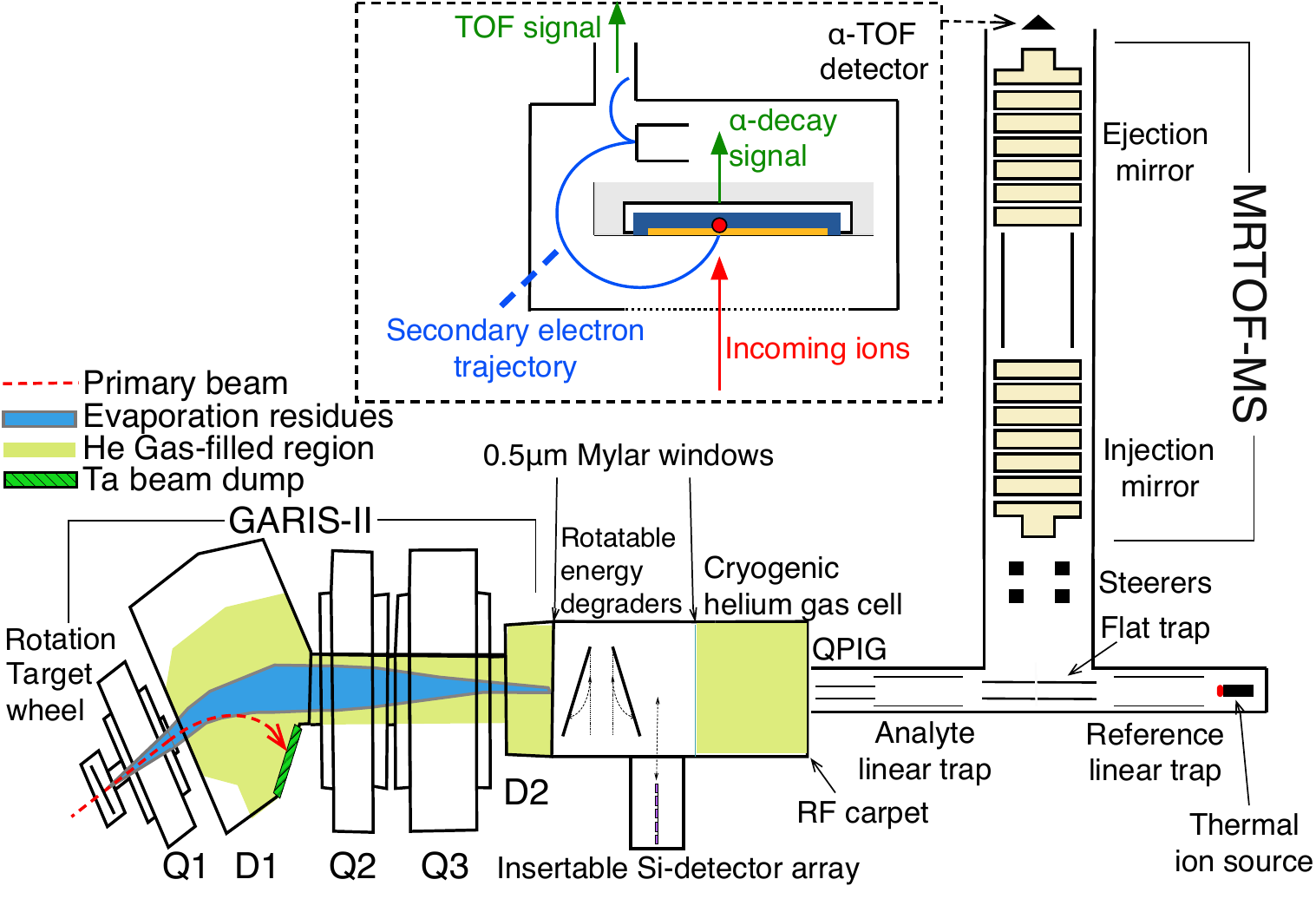}
	\caption{\label{figApparatus} Sketch of the experimental setup.  Fusion-evaporation product ions are separated from the primary beam (projectile) using the gas-filled recoil ion separator GARIS-II.  The ions are stopped and thermalized in the high-purity helium gas cell and subsequently transported and stored in the multiple radio-frequency ion traps before being sent to an MRTOF-MS for mass analysis.  The ion detector at the end of the MRTOF-MS can detect both ion implantation and subsequent $\alpha$-decay.} %The figure taken from reference~\cite{Schury_Db}   \vspace{-1.5 mm}}
\end{figure}

Decay-correlated mass measurements were performed at the SHE-Mass-II facility, jointly operated under the auspices of RIKEN Nishina Center and KEK Wako Nuclear Science Center, within the RIKEN RI Beam Factory.  The experimental setup is shown in Fig.~\ref{figApparatus}. A primary beam of $^{51}$V$^{13+}$ was prepared~\cite{V_beam}, pre-accelerated by the RILAC-II linear accelerator, and injected into the RIKEN \textcolor{black}{Ring} \textcolor{black}{Cyclotron} (RRC) where it was accelerated to \textcolor{black}{306.0~MeV} and impinged upon targets in front of the gas-filled recoil ion separator GARIS-II~\cite{Kaji_garis2}.
\par Sixteen targets of $^{159}$Tb, produced by sputtering onto 3.0-$\mu m$ thick Ti backing foils and having an average thickness of 460-$\mu$g/cm$^{2}$, were mounted on a 16-sector rotating target wheel~\cite{Kaji_rotate}, which rotated at 2000 revolutions per minute during the beam irradiation.  While the peak cross-section for the desired reactions occurs well-below \textcolor{black}{306.0~MeV}, the RRC could not deliver a lower energy beam.  As such, 12.5-$\mu$m Al foil energy degraders were also mounted on the target wheel~\cite{Kaji_double}, upstream of the targets.  By using the energy degraders, the primary beam energy at target center was 219.1~MeV. \textcolor{black}{To pass $^{207}$Ra} the magnetic rigidity of GARIS-II was set to 1.66~T$\cdot$m; the He-gas pressure was set to 71 Pa.  Under this setting, the evaporation residues (ERs) produced in the $^{159}$Tb($^{51}$V, X) reaction were efficiently transported while the unreacted primary beam and other background products were suppressed by GARIS-II. 
\par A secondary beam degrader made from 9.6-$\mu m$-thick Mylar foil was installed in the focal plane chamber of GARIS-II.  The energy-degraded ERs were then stopped and thermalized in a cryogenic helium gas cell, pressurized to 100~mbar room-temperature-equivalent and cooled to 50~K.  The thermalized ions were extracted from the gas cell using a traveling wave radio frequency (RF) carpet~\cite{WADA_RFC}, transferred to an RF ion guide and transported to an RF ion trap suite. After a final cooling process in the ``flat trap"~\cite{Ito_Flat} (see Fig.~\ref{figApparatus}), the ions were orthogonally ejected from the trap and injected into the MRTOF-MS.  In this experiment, the MRTOF-MS was optimized to produce a time focus after 266 laps.
	    
\par If the mean time between ion implantations is too short compared to the $\alpha$-decay half-life, subsequent ion implantation may occur prior to $\alpha$-decay.  Such events can impede decay-correlation and decay spectroscopy. Therefore, for the $\alpha$-decay correlated measurements of $^{207m}$Ra \textcolor{black}{($T_{1/2}$=59~ms)}, the incoming rate of total ion of $^{207}$Ra$^{2+}$ at the MRTOF-MS was limited to about 0.2~cps by reducing the primary beam current to approximately 100~pnA\textcolor{black}{, corresponding to an impinging beam of $6.2\times10^{12}$ particles per second.}  
\par Two measurement runs were performed. The first was six hours in duration, while the second was an hourlong measurement wherein ions made an extra lap in the MRTOF to confirm identifications.  Making measurements at two different numbers of laps ensures that we do not erroneously misattribute TOF peaks from ${e.g.~}$stable molecular ions making an arbitrary number of laps and coincidentally having a TOF similar to our analyte.  In principle, by making correlations between TOF and subsequent $\alpha$-decay, this precaution is not necessary, however it does provide an added layer of confirmation.

\section{Analysis and Results}

\begin{figure}[t]
%	\vspace{2mm}
	\centering
	\includegraphics[scale=0.55]{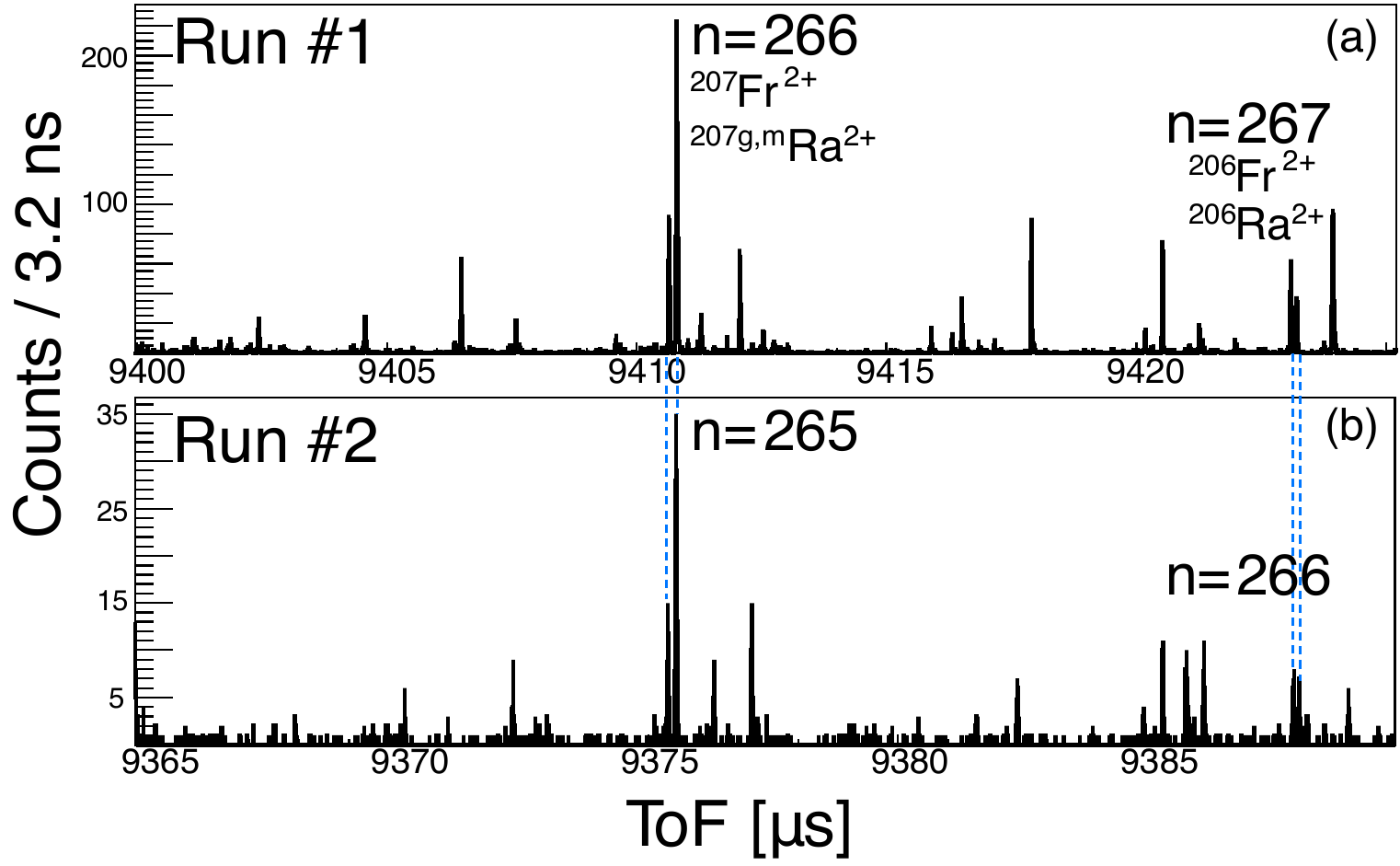}
	\caption{\label{singles_Ra_TOF} The full-range time-of-flight singles spectra for two setting of the MRTOF-MS. (a) The upper panel shows the result of the six hour measurement wherein $A/q$~=~103.5 ions made 267~laps; (b) the lower panel shows the result of the one hour measurement wherein $A/q$~=~103.5 ions made 265~laps. \textcolor{black}{The dashed blue lines indicate the positions of the Fr and Ra isotopes, demonstrating that their relative positions were largely constant under change of lap number between the two runs.} Several peaks, presumed to be stable molecular ions extracted from the gas cell were observed but did not correlate with $\alpha$-decay signals.
	\vspace{-1.5 mm}} 
\end{figure}

Figure~\ref{singles_Ra_TOF} shows the full-range TOF singles spectra for the two runs.  In Run \#1, the MRTOF-MS timing was configured such that $A/q$~=~103.5 ions would make 266~laps in the MRTOF and $A/q$~=~103 ions would make 267~laps; care was taken to ensure ions of both $A/q$ were not effected by the ejection-side mirror switching~\cite{Schury_wide2}. Among the $A/q$~=~103.5 ions, $^{207}$Ra$^{2+}$ and $^{207}$Fr$^{2+}$ were identified, while $^{206}$Ra$^{2+}$ and $^{206}$Fr$^{2+}$ were identified among the $A/q$~=~103 ions.  To exclude misidentification of these ions, in Run \#2 the ejection mirror timing was adjusted such that $A/q$~=~103.5 ions made 265~laps and $A/q$~=~103 ions made 266~laps. As the same identifications were made in this hourlong cross-check measurement, we can be confident of the identifications.
\textcolor{black}{The lack of particularly mass selective elements between the gas cell and MRTOF automatically allows for the simultaneous study of multiple $A/q$ chains. The tune of the MRTOF used in this work resulted in $A/q=103, 103.5$ chains having flight lengths differing by one lap, and their relative separation changes only slightly with small variations in the lap number.  Other isobaric chains, however, will generally exhibit larger changes in their relative position in the time-of-flight spectrum under small variations in the lap number.}
\par The times-of-flight were determined from TOF spectral peaks by fitting using an asymmetric Gaussian-hybrid function~\cite{Marco}:
	\begin{eqnarray}
	f(t)=\left\{ \begin{array}{ll}
	A e^{\delta_{\rm L}(2t-2t_{\rm c}+\delta_{\rm L})/2\sigma^{2}} & ( t\leq t_{\rm L} ), \\
	A e^{-(t-t_{\rm c})^{2}/2\sigma^{2}} & ( t_{\rm L} ~\textless t ~\textless t_{\rm R}),\\
	A e^{\delta_{\rm R}(-2t+2t_{\rm c}+\delta_{\rm R})/2\sigma^{2}} & ( t\geq t_{\rm R} ),  
	\label{eq:Hybrid-two}
	\end{array} \right.
	\end{eqnarray}
where $A$ is the Gaussian peak height, $t_{\rm c}$ is the Gaussian centroid ({\emph i.e.} the time-of-flight) and $\sigma$ is the standard deviation.  The fit function smoothly transitions from Gaussian to exponential on both sides of the peak, with the transition point defined to be $t_{\rm L}=t_{\rm c}-\delta_{\rm L}$ on the left side and $t_{\rm R}=t_{\rm c}+\delta_{\rm R}$ on the right side. \par

The data in Fig.~\ref{singles_Ra_TOF} has been drift-corrected using $^{85}$Rb$^{+}$ ions from a thermal ion source.  These reference ions were measured concomitantly~\cite{Schury_concomi} with the analyte ions shown in Fig.~\ref{singles_Ra_TOF}.  The time-of-flight spectra were divided into subsets of 7.5~s duration, each was fitted using Eq.~(\ref{eq:Hybrid-two}), and then the times-of-flight of every ion (reference and analyte) in each subset were adjusted to compensate for drift.  
\par The masses $m$ of ions with charge $q$ were determined using the single-reference method~\cite{Kimura, Ito_Li}
	\begin{equation}
	m=\frac{q}{q_{\rm ref}}\rho^2 m_{\rm ref}=\frac{q}{q_{\rm ref}}\biggl(\frac{t-t_{0}}{t_{\rm ref}-t_{0}}\biggl )^{2}m_{\rm ref},	
	\label{eq:ratio}
	\end{equation}
	where $\rho$ is the time-of-flight ratio between the analyte and reference, and $t_{0}$ is the delay between the time-to-digital converter start signal and the ejection from the preparation ion trap which sends ions to the MRTOF-MS. When analyte and reference ions are isobaric, the contribution of $t_0$ becomes negligible and can be ignored. 
The analyte ions were identified from their time-of-flight ratios with the $^{85}$Rb$^{+}$ ions, using Eq.~(\ref{eq:ratio}) with $t_0$~=~40(4)~ns. In this work $^{206,207}$Fr$^{2+}$ ions were used as isobaric references in the precise determination of the masses of $^{206,207g,m}$Ra.

\par
\begin{figure}[t] 
%	\vspace{2mm}
	\centering
	\includegraphics[scale=0.44]{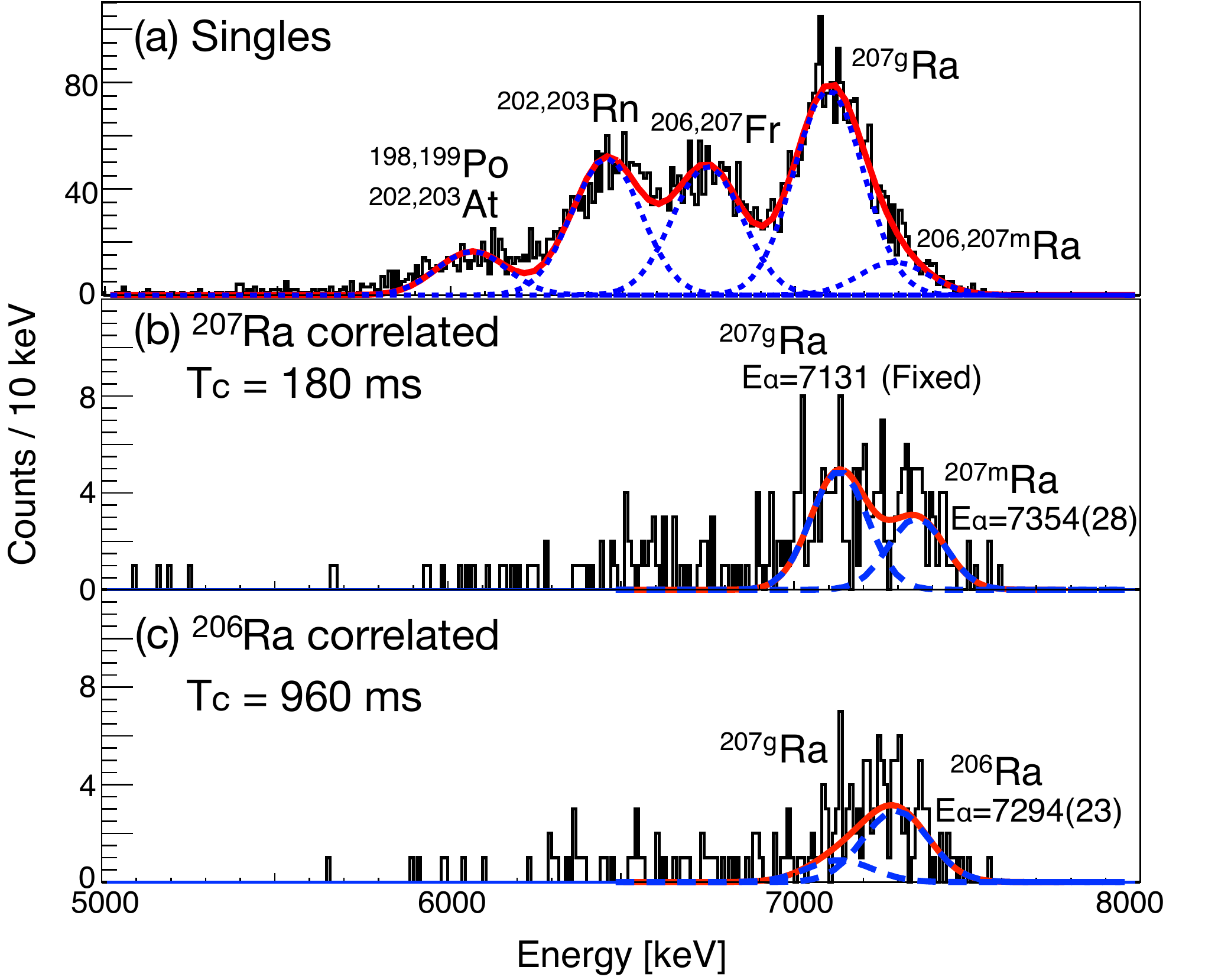}
	\caption{\label{Singles_Ra_Alpha} The alpha spectra obtained by $\alpha$-TOF detector during the six hour measurement. (a) Alpha singles spectrum. (b) The TOF-correlated $\alpha$-decay spectrum gated on $^{207}$Ra$^{2+}$. (c) The TOF-correlated $\alpha$-decay spectrum gated on $^{206}$Ra$^{2+}$. \vspace{-1.5 mm}} 
\end{figure}

 The $\alpha$-singles spectrum measured during Run\#1 is shown in Fig~\ref{Singles_Ra_Alpha} (a).  Directly transported $^{207}$Fr, $^{207g/m}$Ra, $^{206}$Fr, and $^{206}$Ra were observed, along with their Rn, Po, and At isotope decay products.  Figures.~\ref{Singles_Ra_Alpha} (b) and (c) show the TOF-correlated $\alpha$-decay spectra for $^{207}$Ra$^{2+}$ and $^{206}$Ra$^{2+}$, respectively. \par
\textcolor{black}{The probability of an accidental coincidence is calculated from the ratio of the total coincidence time gate to the total measurement time. In the analysis of $^{207}$Ra shown in Fig.~\ref{Singles_Ra_Alpha}~(b), we selected 180~ms as the coincidence time ($T_{\rm c}$), which corresponds to three half-lives of $^{207m}$Ra. According to this coincidence time and the counting of time-of-flight signals, we estimated that 1.8\% of the $^{207g}$Ra produced accidental coincidences.  Consequently, the decay-coincidence gated ToF spectrum is contaminated with 8.6\% of the counts being $^{207g}$Ra$^{2+}$.  Nonetheless, it can be seen that the isomeric component is greatly enhanced compared to singles spectrum.}
% the accidental coincidence probability induced from $^{207g}$Ra is estimated to be 1.8\%. In addition, a further 8.6\% of $^{207g}$Ra decays in the coincidence time gate and appears as a contaminant in the isomeric state discrimination. However, it can be seen that the isomeric component is enhanced compare to singles spectrum. In Fig.~\ref{Singles_Ra_Alpha}~(c) we selected the coincidence time of $T_{\rm c}$=~960 ms.  
\par \textcolor{black}{In the analysis of $^{206}$Ra, shown in Fig.~\ref{Singles_Ra_Alpha}~(c), we selected a coincidence time of $T_{\rm c}$=~960 ms.  The accidental coincidence rate induced by $^{207g}$Ra, $^{206,207}$Fr, etc. was calculated to be 1.5\%.  In this case there are no isomeric states to hinder the correlation analysis.}
\par
	\textcolor{black}{The decay energies of $^{207m}$Ra and $^{206}$Ra were determined to be 7.354(28)~MeV, and 7.294(23)~MeV, respectively, by application of least-square fitting of the TOF-correlated $\alpha$-decay spectra. To provide a calibration in the determination of absolute decay energy, the alpha-decay energy of $^{207g}$Ra was fixed to the literature value of 7.131~MeV~\cite{NDS_207}. The intensity of $^{207g}$Ra used in the fitting process was fixed by the calculated number from the $\alpha$-decay singles spectrum.} These results are in agreement with literature values derived from precise $\alpha$-decay measurements~\cite{NDS_206, NDS_207}.


\subsection{$^{206}$Ra}
\begin{figure}[tb]
%	\vspace{2mm}
	\centering
	\includegraphics[scale=0.45]{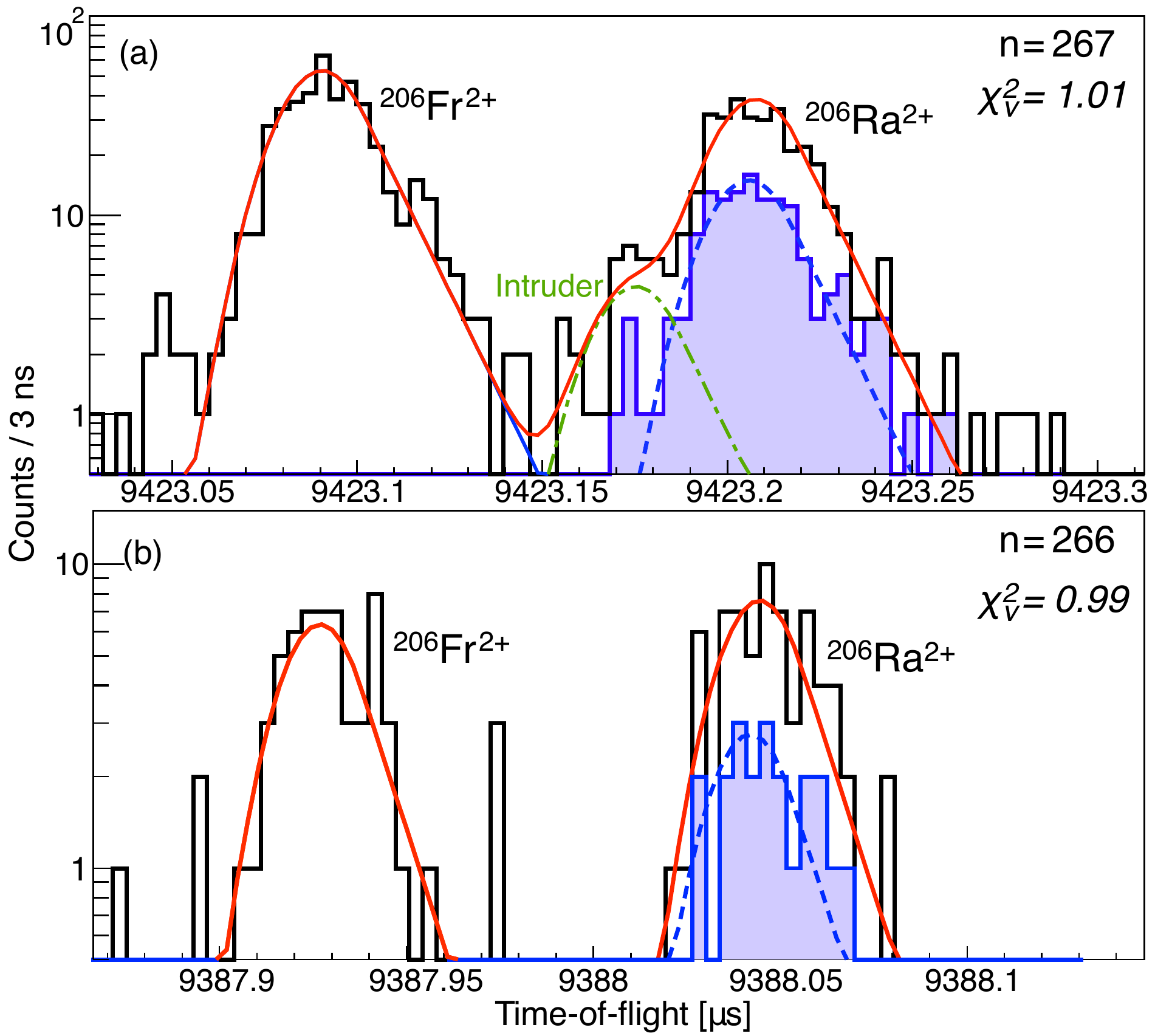}
	\caption{\label{206Ra_run} Time-of-flight spectra from the two runs: (a) Run \#1 with 267 laps, (b) Run \#2 with 266 laps.  The TOF singles spectra are drawn in black, while the $^{206}$Ra $\alpha$-decay correlated spectra are drawn in blue filled. The coincidence period of the gate was 1~s, corresponding to approximately four half-lives. \textcolor{black}{The reduced $\chi^{2}$-value ($\chi^{2}_{\nu}$) for each of the fits are indicated in below the number of laps.} \vspace{-1.5 mm}} 
\end{figure}

Figure~\ref{206Ra_run} shows the time-of-flight spectra measured in the two runs, centered on the ions $^{206}$Fr$^{2+}$ and $^{206}$Ra$^{2+}$. The TOF singles spectra are shown in black, while the $^{206}$Ra $\alpha$-decay correlated TOF spectra are shown in blue.  A spurious ion species, presumably a stable molecular ion, can be seen in the tail of the $^{206}$Ra$^{2+}$ case of 267~laps, but it is suppressed by the decay correlation.
\par Within a region of $\pm$50~ns of the center of the $^{206}$Ra$^{2+}$ spectral peaks there were 162 decay-correlated TOF events \textcolor{black}{using $T_{\rm c}$=~960~ms}, out of 327 TOF singles events. The fraction of correlated events, 49(5)\%, agrees well with the previously reported $\alpha$-TOF efficiency~\cite{Niwase}.
\par The isobaric molecular ions should all have the same peak shape, and therefore we simultaneously fit the two spectra (singles and $\alpha$-decay correlated) for each run with the width and exponential tail parameters conserved across all peaks.  While the spurious intruder ion may exhibit a slightly different shape, its low intensity precludes any minor shape difference resulting in significant biasing of the fit parameters.  
\par A more pressing issue, if we wish to use isobaric referencing, is the isomerism of $^{206}$Fr.  Were all three states delivered, the second isomer ($E_\textrm{ex}$~=~730~keV) would be resolvable in the TOF spectra with the mass resolving power of this experimental conditions ($R_{m}\approx $178,000) but the first isomer ($E_\textrm{ex}$~=~190~keV) would be unresolvable.  However, were $^{206}$Fr to be delivered as an admixture of ground state and first isomer with comparable intensities, the peak width would be noticeably broadened. As the $^{206}$Fr$^{2+}$ and $^{206}$Ra$^{2+}$ peaks are well-reproduced with a shared peak width, we can presume the $^{206}$Fr$^{2+}$ peak is dominated by either the ground state or first isomer. In an experiment conducted at ISOLDE, the population of $^{206g}$Fr produced by the spallation reaction of UC$_{x}$ was about two times larger than that of $^{206m1}$Fr~\cite{Laser_Fr}, so we presume $^{206g}$Fr$^{2+}$ to be the highly dominant state in our analysis.

%Black histogram indicates the singles spectrum and blue histogram indicates the decay correlated time-of-flight histogram. The reported half-life of $^{206}$Ra is 240(20)~ms, hence the coincidence time was set to 1~s corresponding to about 4 half-life period.

% At 267 laps, a spurious spectral peak (labeled ``Bump" in Fig.~\ref{206Ra_run} (a)), which presumed to be a molecular ion having a different lap numbers, was found on the left side of $^{206}$Ra$^{2+}$. Therefore, the fitting of singles spectrum include a bump structure. In order to preclude any mass-dependent systematic errors, $^{206}$Fr$^{2+}$ used as an isobaric reference~\cite{Kimura}. The fitting was done for $^{206}$Fr$^{2+}$ and $^{206}$Ra$^{2+}$ and bump structures at once to determine the shape of the fitting function and the time-of-flight ratio and fitting errors at the same time.

%	where $t_{0}$ is the delay between the TDC (time-to-digital converter) start signal and the ejection from the preparation ion trap of MRTOF-MS.
\par Table~\ref{tab:Result_206} summarizes the mass analysis results for $^{206}$Ra$^{2+}$.  The weighted average mass excess of $^{206}$Ra determined from the TOF singles spectra was $M\!E$~=~3548(50)~keV/c$^2$, while the decay-correlated data yielded $M\!E$~=~3460(92)~keV/c$^2$.  Both of these values are in agreement with the $M\!E_{\rm Lit}$~=~3566(18) keV/c$^2$ literature value~\cite{AME2020}, confirming the device performance.
\par
\begin{figure}[tb]
%	\vspace{2mm}
	\centering
	\includegraphics[scale=0.49]{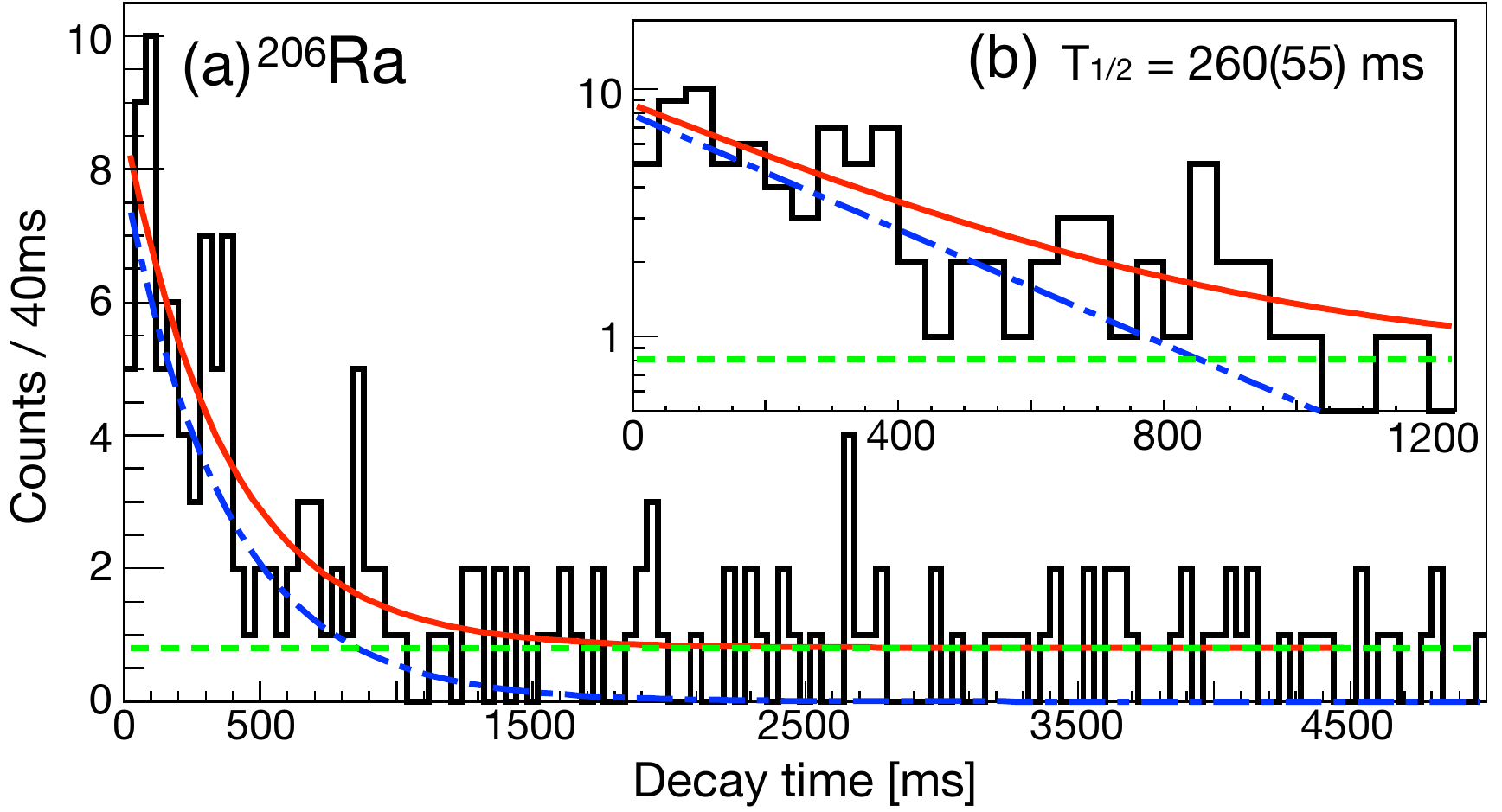}
	\caption{\label{206Ra_decaytime} 
	Decay time spectrum as a histogram of the time interval between the ion implantation signal used to determine the time-of-flight and the subsequent $\alpha$-decay signal of appropriate energy: (a) full range up to 5~s, (b) the first 1.2~s. The half-life of $^{206}$Ra was determined to be 260(55)~ms by fitting this data as an exponential decay with constant background (solid red curve).
%	Decay time distribution of decay-tagged $^{206}$Ra events. Red line indicate the fitting exponential function for evaluate to half-life.  
\vspace{-1.5 mm}} 
\end{figure}

 In addition to mass measurements, the $\alpha$-TOF detector allows for determination of half-lives.  Figure~\ref{206Ra_decaytime} provides a histogram of the time between ion implantation and subsequent $\alpha$-decay, gated on the 7250\textcolor{black}{$\pm$250}~keV alpha-decay energy characteristic of $^{206}$Ra in Run\#1.  By fitting the histogram data to an exponential decay with constant background a half-life for $^{206}$Ra of $T_{1/2}$~=~260(55)~ms was determined. From Run\#2, totally 21 decay correlated events were obtained. These half-lives were determined to be $T_{1/2}$~=~360(100)~ms from the mean value, as the maximum likelihood value~\cite{Schmidt}. The weighted average of Run \#1 and Run \#2 was calculated and the half-life of $^{206}$Ra was determined to be $T_{1/2}$~=~283(48)~ms, which is in good agreement with the $T_{1/2}$~=~240(20)~ms literature value~\cite{NNDC}. From the experimental results and literature values, a new global half-life of $^{206}$Ra, $T_{1/2}$~=~248(18)~ms, is obtained.
%\par The results of the fitting to 267 laps TOF singles spectral peaks of $^{206}$Fr$^{2+}$ and $^{206}$Ra$^{2+}$ yielded a TOF ratio $\rho$ = 1.000012462(142), which corresponding to the mass excess of 3540(54)~keV. Repeating the fitting procedure using decay-correlated spectrum yielded a TOF ratio $\rho$ =1.000012256(270), correspond to the mass excess of 3461(103)~keV. Additionally, a half-life analysis was performed using the TOF gated decay time spectrum. The decay time of all correlated events are compiled in Fig.~\ref{206Ra_decaytime} as a histogram. The exponential fitting gives the half-life of $T_{1/2}$=260(55)~ms. This value is consistent with the literature value of $T_{1/2}$=240(20)~ms~\cite{NNDC}.\par
%	The same mass analysis was performed for the 266 lap spectrum for confirmation. The time-of-flight ratios in this case were evaluated to be $\rho$=1.000012604(340) from singles TOF spectrum and $\rho$=1.000012249(556) from decay-correlated time-of-flight signal, which corresponding to mass excess of 3594(130)~keV and 3458(213)~keV, respectively.\par
%Table~\ref{tab:Result_206} summarizes the mass analysis results for $^{206}$Ra$^{2+}$. For the results of 267 laps with the higher statistics, the deviation from the literature value of mass excess, 3566(18)~keV, was $-$26(57)~keV in singles events and $-$105(104)~keV in decay-correlated events. In both cases, the result agrees with literature values within statistical error, indicating that accurate masses can be derived from decay-correlated TOF spectra.

\begin{table}[tb]
\caption{
The measured time-of-flight ratio $\rho$ between $^{206}$Ra$^{2+}$ ion and the reference ion ($^{206}$Fr$^{2+}$), mass excess ($M\!E$) and half-life of of $^{206}$Ra from TOF singles and decay-correlated TOF data. The mass deviations ($\Delta M\!E$) are comparisons with the literature value $M\!E_{\rm  Lit}$ from AME2020~\cite{AME2020}.
\label{tab:Result_206}
}
\scalebox{0.8}{
\begin{tabular}[t]{ccccc} \hline \hline
laps & $\rho$ & $M\!E_{\rm{EXP}}$ [keV] & $\Delta M\!E$ [keV] & $T_{1/2}$ [ms]  \\ 
\hline 
Singles \\
\hline
267 & 1.000012462(142) & 3540(54) & $-$26(57) & \\
266 & 1.000012604(340) & 3594(130) & 28(131)  \\
\hline
Correlated \\
\hline
267 & 1.000012256(270) & 3461(103) & $-$105(104) & 260(55)  \\  
266 & 1.000012249(556) & 3458(213) & $-$107(214)  & 360(100) \\ 
\hline
Weighted Averages \\
\hline
Singles & 1.000012483(131) & 3548(50) & -18(53) \\
Correlated & 1.000012255(243) & 3460(93) & -103(95) & 283(48) \\
\hline\hline
\end{tabular}
}
\end{table}

\subsection{$^{207}$Ra}
\begin{figure}[tb]
%	\vspace{2mm}
	\centering
	\includegraphics[scale=0.49]{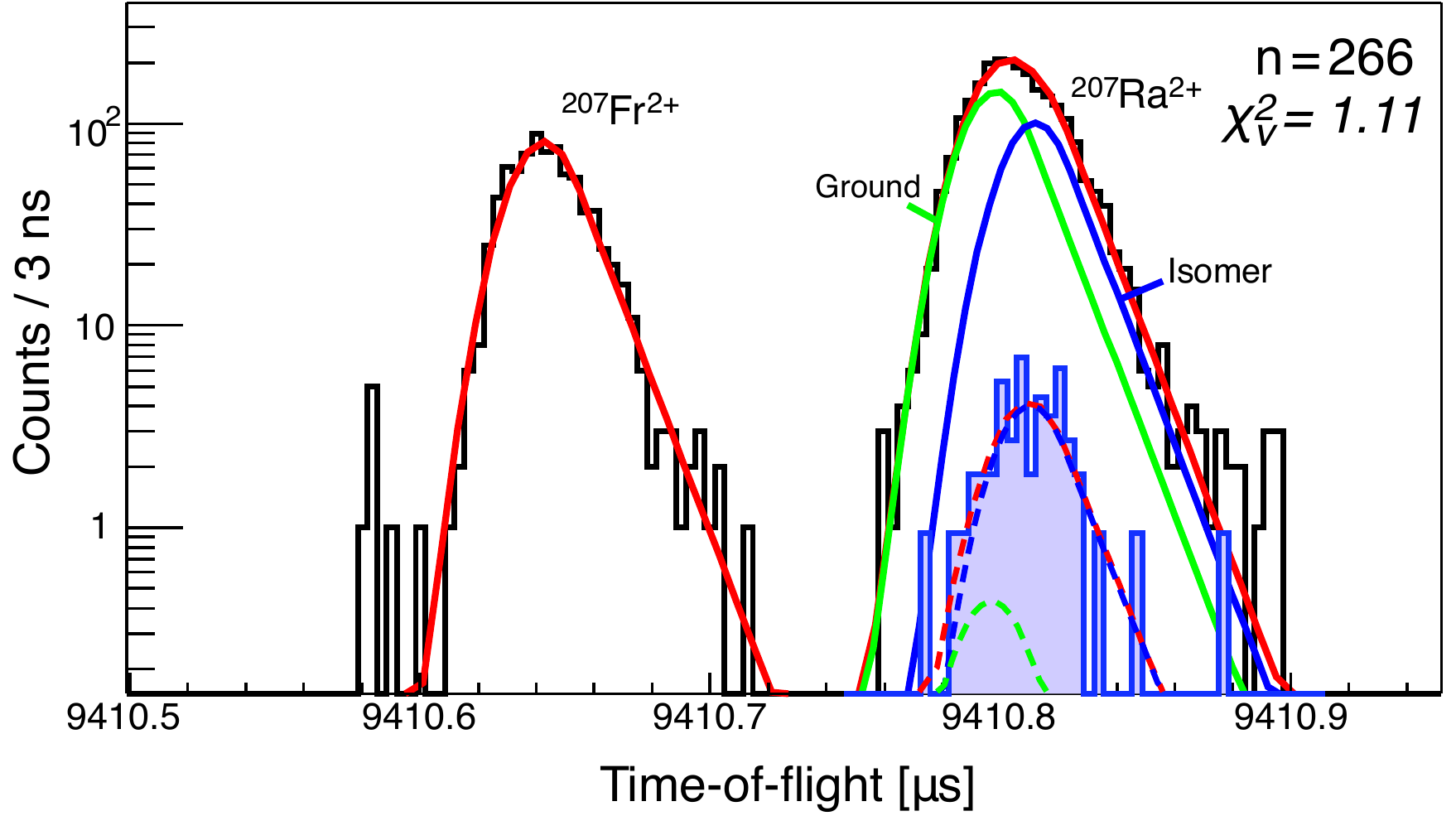}
	\caption{\label{207Ra_TOF_180} The time-of-flight spectrum around $^{207}$Ra$^{2+}$ region along with the fitting results. The black and blue filled histograms are the singles and decay-correlated TOF events, respectively. The solid lines indicate the fitting curves of the singles histogram for ground state (green), isomeric state (blue), and summing of both states (red). The dashed lines similarly represent the fitting curves for the decay-correlated TOF spectrum.  \vspace{-1.5 mm}} 
\end{figure}

A similar set of analyses was performed for $^{207}$Ra.  Figure~\ref{207Ra_TOF_180} shows the time-of-flight spectra accumulated in Run \#1, with the TOF singles histogram in black and the $^{207m}$Ra $\alpha$-decay correlated TOF histogram in blue; due to the incoming rate of $^{207g}$Ra$^{2+}$ ($T_{1/2}$=1.38 s) higher than the decay rate, it was not possible to perform decay correlations for the ground state.  To minimize incidental correlations between $^{207g}$Ra$^{2+}$ TOF and $^{207g}$Ra $\alpha$-decay, the energy gate had a lower limit of 7.32~MeV (2$\sigma_E$ from the $^{207g}$Ra $\alpha$-decay energy) resulting in slight reduction in the efficiency.  A coincidence time gate of $T_{c} <$~180~ms, corresponding to approximately three half-lives of $^{207m}$Ra, was used.  This is shorter than the typical four half-lives, again to suppress the amount incidental coincidence with $^{207g}$Ra decays.  Despite these efforts we find that 14\% of the counts in the decay-correlated ToF spectra (blue histogram in Fig.~\ref{207Ra_TOF_180}) are derived from $^{207g}$Ra$^{2+}$ as noted by the dashed green curve in Fig. 6.
%Figure~\ref{207Ra_TOF_180} shows the time-of-flight spectra near the $^{207}$Fr$^{2+}$, $^{207}$Ra$^{2+}$ making 266 laps (Run~\#1). In the spectrum, we show the singles events overlapped with the decay-correlated events of $^{207}$Ra. As the incoming rate of $^{207g}$Ra ($T_{1/2}$=1.38 s) was slightly excessive compared the decay rate, we could not perform decay-correlated time-of-flight spectroscopy for that state. Therefore, we focused on $^{207m}$Ra, which has been reported to have a short half-life, $T_{1/2}=$59 ms. To avoid contamination by ground state, the energy gate of 7.32$\geq$MeV was used, corresponds to 2$\sigma$ from the center of the alpha decay energy of the ground state of $^{207}$Ra. The coincidence time gate ($T_{c}$) was set to $T_{c} <$180 ms, which corresponds to three half-life period of $^{207m}$Ra.\par
The two states of $^{207}$Ra could not be resolved in the singles time-of-flight spectrum alone; the fitting of the two components did not converge, although the peak width indicated multiple components were present.

\par Similar to the case of $^{206}$Ra, the spectral peaks were fit such that the TOF singles spectral peaks of $^{207g,m}$Ra$^{2+}$ and $^{207}$Fr$^{2+}$, the decay-correlated TOF spectral peak of $^{207m}$Ra$^{2+}$, and the incidentally correlated $^{207g}$Ra$^{2+}$ \textcolor{black}{had mutually-fixed values for the peak-width and exponential tail parameters}; the fraction of $^{207g}$Ra$^{2+}$ ions in the fit of the decay-correlated spectrum was fixed to 14\%. Using the position of the isomeric state determined from decay-correlated events as  a fixed parameter when fitting the TOF singles spectrum, the two-component fit reliably converged and the times-of-flight for the two states could be ascertained.

% While the two states of $^{207}$Ra could not be resolved in the time-of-flight spectrum, their separation was sufficient to noticeably broaden the conjoined spectral peak of $^{207g,m}$Ra$^{2+}$, providing sufficient separation to allow the two-component fits to converge reliably.

% The fitting of the TOF spectrum was done simultaneously $^{207g/m}$Ra$^{2+}$ TOF singles, decay-correlated $^{207m}$Ra$^{2+}$ events, and $^{207}$Fr$^{2+}$ TOF singles which served as the isobaric mass reference in one function, to determine the fitting shape and the peak position of the each states of $^{207}$Ra. By sharing parameters between the singles $^{207}$Ra and the decay-correlated $^{207}$Ra, the position of the isomer is determined with high accuracy from the decay-correlated event as fitting with high statistics singles event. The decay-correlated $^{207}$Ra spectrum is mostly attributed to the isomeric state, however a 14\% ground state contribution is estimated from the energy and decay time gates. Therefore, the fitting took into account this mixture of the ground state. \par

\par The masses of $^{207g,m}$Ra$^{2+}$ were determined from the ratio of their times-of-flight with that of $^{207}$Fr$^{2+}$ using Eq.~\ref{eq:ratio}.  Decay-correlated spectra were produced using a coincidence time of 180~ms, as previously discussed. From these measurements, the time-of-flight ratio of $^{207g,m}$Ra$^{2+}$ were measured $\rho_{g}$~=~1.000016550(38) and $\rho_{m}$~=~1.000017983(109) with reference species $^{207}$Fr$^{2+}$. The mass excess of $^{207g}$Ra and excitation energy of $^{207m}$Ra are determined to be $M\!E~=$~3538(15)~keV/c$^2$ and $E_{\rm ex}$~=~552(42)~keV respectively, which consistent with previous indirectly determined values, $M\!E_{\rm Lit}~=$~3510(60)~keV/c$^2$ and $E_{\rm ex,Lit}$~=~554(15)~keV, based on $\alpha$-decay spectroscopy~\cite{AME2020,NDS_207}.
%The results were analyzed with a coincidence time of 180~ms, a time-of-flight ratio of $\rho$=1.000016550(38) with reference species $^{207}$Fr$^{2+}$ and a mass excess of $^{207g}$Ra and excitation energy of $^{207m}$Ra are 3538(15)~keV and 552(42)~keV, respectively. These values are consistent with those determined indirectly by $\alpha$-decay spectroscopy. 

\par
\begin{figure}[t]
%	\vspace{2mm}
	\centering
	\includegraphics[scale=0.49]{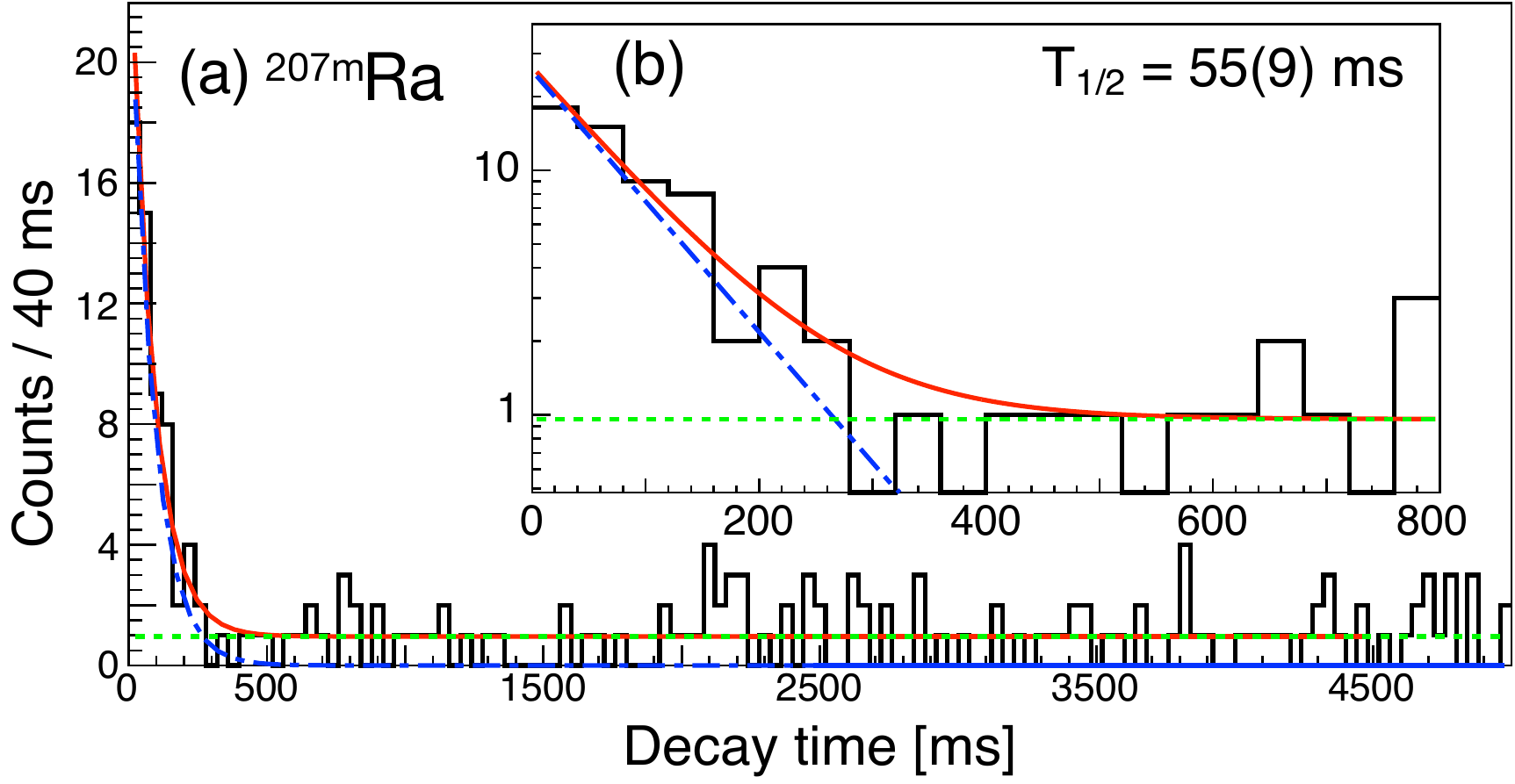}
	\caption{\label{207Ra_decaytime} The decay time distribution of $^{207m}$Ra, determined with an energy gate of E$_{\alpha}\geq$~7.32~MeV. The solid red lines indicate the fitted decay curve of $^{207m}$Ra, while the dashed green lines indicate the constant background level and the long dashed dotted blue line represents the background-subtracted decay curve. (a) Full range spectrum up to 5 seconds, linear scale. (b) Focus on the first 500 ms, logarithmic scale.    \vspace{-1.5 mm}} 
\end{figure}

 Additionally, the half-life of $^{207m}$Ra was determined using the decay-correlated TOF events. The decay time distribution, determined with an energy gate of E$_{\alpha}\geq$~7.32~MeV applied to the data of Run \#1, is shown in Fig.~\ref{207Ra_decaytime}. The fitting results show that the half-life of $^{207m}$Ra is 55(9)~ms in agreement with the literature value of 59(4)~ms.
\textcolor{black}{From Run \#2, we obtained only one decay-correlated event of $^{207m}$Ra. While these are low statistics, it is consistent with the number assumed from the production and alpha branching ratios (discussed later). There are 317 TOF singles events of $^{207m}$Ra, and the expected correlated events based on energy gate width and detection efficiency would be 2.6 events. When considering the Poisson distribution, the probability of getting less than one event when 2.6 events are expected is 27\%, which is statistically reasonable. Unfortunately, however, a similar analysis to that of Run \#1 data is not possible. Therefore, the data obtained from Run \#2 were used primarily to confirm the identify of $^{207}$Fr$^{2+}$ and $^{207}$Ra$^{2+}$ based on their unchanging relative times-of-flight. }

\par The first study of $^{207m}$Ra was reported in He{\ss}berger \it{et al.}~\rm{\cite{Hess}.} They estimated the production ratio of $^{207g}$Ra/$^{207m}$Ra to be 0.75/0.25 in the reaction system of $^{58}$Fe+$^{154}$Sm, and that the $\alpha$-branching ratio of $^{207m}$Ra was $b_{\alpha}<$~0.25. Subsequent experiments by Leino \it{et al.}~\rm{\cite{Leino},} have also supported this $\alpha$-decay branching ratio. 

\par Since we can unambiguously determine the ratio of ground state to isomer from the TOF singles spectrum, it is also possible to derive the alpha-decay branching ratio and partial half-life of $^{207m}$Ra using the $\alpha$-TOF detector. To determine the $\alpha$-decay branching ratio we must determine the number of $\alpha$-decays from $^{207m}$Ra which occurred at the detector and the total number $^{207m}$Ra atoms deposited on the detector during the experiment. 
\par  While we cannot  fully resolve the spectral peaks from $^{206}$Ra and $^{207m}$Ra in the alpha singles spectra of Fig.~\ref{Singles_Ra_Alpha} (a), the fitting results indicate the total number of $\alpha$-decays $N_{sum}$~=~$N_{206\rm{Ra}}$ + $N_{207m\rm{Ra}}$~=~305~$\pm$~21.4 from $^{206}$Ra and $^{207m}$Ra were detected.  The TOF-correlated $\alpha$-decay spectrum gated on $^{206}$Ra$^{2+}$ consisted of $N_{206\rm{Ra}}$~=~162~$\pm$~12.7.  Thus, after correcting $N_{sum}$ for detector efficiency, we conclude that $N^\prime_{207m\rm{Ra}}$~=~245~$\pm$~49.5 $\alpha$-decays from $^{207m}$Ra occurred on the $\alpha$-TOF detector in the course of the measurement.
\par The total number of $^{207m}$Ra deposited on the detector can be determined from the TOF singles spectra (Fig.~\ref{207Ra_TOF_180}).  The two-component fit of the $^{207g,m}$Ra$^{2+}$ spectral peak indicated a ratio of 0.60(2)/0.40(2) for the ground state to isomer yield, similar to the yield ratio observed in the $^{58}$Fe+$^{154}$Sm reaction system.  After correcting for the TOF detector efficiency, we determined that 945 $^{207g,m}$Ra$^{2+}$ ions implanted upon the detector during the measurement.  Thus, the $\alpha$-decay branching ratio of $^{207m}$Ra could be determined to be 26(20)\%, in agreement with the value of 25\% previously reported from $\alpha$-decay spectroscopy~\cite{Hess,Leino}.

\section{Discussion}

\par
\begin{figure}[t]
%	\vspace{2mm}
	\centering
	\includegraphics[scale=0.55]{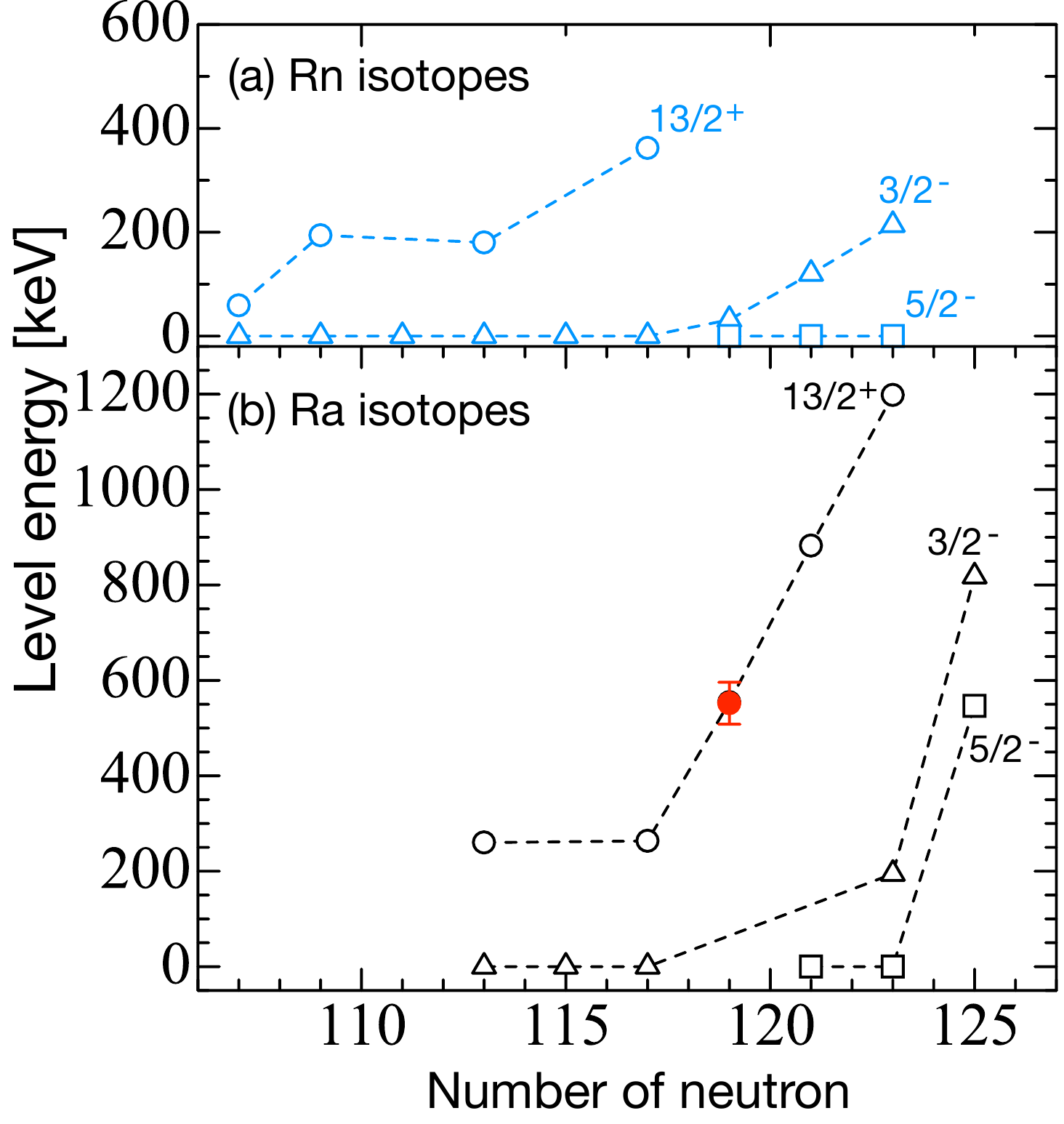}
	\caption{The single particle level diagram for odd Ra and Rn nuclei. The closed circle indicates the data from this experiment. \label{Rn_Ra_single}   \vspace{-1.5 mm}} 
\end{figure}

 The spin-parity of $^{207m}$Ra has tentatively assigned to $J^{\pi}$~=~13/2$^{+}$, based on the systematics of the neighboring nuclei~\cite{NDS_207}. The single particle level diagrams for odd Ra and Rn nuclides are shown in Fig.~\ref{Rn_Ra_single}. The solid red circle in Fig.~\ref{Rn_Ra_single} is the excitation energy of $^{207m}$Ra obtained from our decay-correlated mass analysis. It agrees with the prior study from $\alpha$-decay spectroscopy. The systematics continue to suggest that spin-parity is $J^{\pi}$~=~13/2$^{+}$. The spin-parity of $^{207g}$Ra is assigned to either $J^{\pi}$~=~3/2$^{-}$ or 5/2$^{-}$ based on systematics. If $J_{\pi}$~=~3/2$^{-}$ then the configuration is (($\pi$h$_{9/2}$)$^{+6}_{0+}(\nu$p$_{3/2})^{-1})$ while if $J_{\pi}$~=~5/2$^{-}$ then the configuration would be (($\pi$h$_{9/2}$)$^{+6}_{0+}(\nu$f$_{5/2})^{-1})$~\cite{NDS_207}. However, the spin-parity of $^{207g}$Ra cannot be determined from this experiment.

\par From our results, the reduced alpha width of $^{207m}$Ra can be evaluated to be \textcolor{black}{$\delta^{2}$~=~$40^{+68}_{-34}$~keV.} according to the Rasmussen prescription~\cite{Rasmussen}. Table~\ref{tab:Reduced_Ra} describes the reduced alpha width of 13/2$^{+}$ and 3/2$^{-}$ state in the neighboring nuclei $^{203}$Ra and $^{205}$Ra. Prior to our study, the reduced alpha width of $^{203}$Ra and $^{205}$Ra were reported to be around 60~keV for both states~\cite{Kalaninova}. $^{207m}$Ra is suggested to be 13/2$^{+}$ not only in terms of single particle level systematics, but also from the reduced alpha width systematics. The reduced width is consistent with that of the neighboring nuclides, indicating that the $\alpha$-decay of $^{207m}$Ra is not forbidden.

\begin{table}[htb]
\begin{center}
\caption{Summary of the spin-parity, decay properties and reduced alpha width ($\delta^{2}$) of odd $^{203-207}$Ra isotopes. $b_{\alpha}$ is the branching ratio of $\alpha$-decay.\label{tab:Reduced_Ra}}
\begin{tabular}[t]{c c c c c c c}
\hline \hline
Isotope & $J^{\pi}$ & $E_{\alpha}$ [keV] & $T_{1/2}$ [ms] & $b_{\alpha}$ & $\delta ^{2}$ [keV] & Ref.  \\ \hline \hline
$^{203g}$Ra & 3/2$^{-}$ & 7575(10) & $50^{+40}_{-15}$ & 1.0 & $45^{+37}_{-14}$ & \cite{Kalaninova} \\
$^{203m}$Ra & 13/2$^{+}$ & 7607(8) & $37^{+37}_{-12}$ & 1.0 & $48^{+48}_{-16}$ & \cite{Kalaninova} \\ \hline
$^{205g}$Ra & 3/2$^{-}$ & 7340(20) & $210^{+60}_{-40}$ & 1.0 & $50^{+24}_{-17}$ & \cite{NNDC} \\
$^{205m}$Ra & 13/2$^{+}$ & 7370(20) & $170^{+60}_{-40}$ & 1.0 & $48^{+27}_{-18}$ & \cite{NNDC} \\ \hline
$^{207g}$Ra & 3/2$^{-}$\footnote{tentatively assigned~3/2$^{-}$ or 5/2$^{-}$.} & 7131(5) & $1380^{+220}_{-110}$ & 0.86 & $37^{+5}_{-6}$ & \cite{NNDC} \\
$^{207m}$Ra & 13/2$^{+}$ & \textcolor{black}{7354(28)} & $55(9)$ & 0.26(20) & \textcolor{black}{$43^{+68}_{-34}$} & this work \\ \hline \hline
\end{tabular}
\end{center}
\end{table}

\section{Summary}

This work has shown that the correlation measurements of mass and decay properties by use of an MRTOF-MS equipped with the $\alpha$-TOF detector, proving that not only the unique determination of the excitation energy of an isomeric state, but also its branching ratio and partial half-life can be derived simultaneously. The results are in agreement with the historical $\alpha$-spectroscopy measurements, and successfully demonstrated the value of simultaneous mass and decay spectroscopy with MRTOF-MS equipped with the $\alpha$-TOF detector. In particular, the mass of $^{207g}$Ra and the excitation energy of $^{207m}$Ra were directly measured for the first time. The nuclear structure could be discussed based on analyses of the decay-correlated time-of-flight spectrum. The exploration of this technique is expected to contribute to further investigations the nuclear structure of heavy and superheavy nuclides.\par
Recently, we have achieved a mass resolving power $R_{m}\approx$~700,000 for our MRTOF system. Additionally, the improvement of the energy resolution of the $\alpha$-TOF detector is currently underway. 
The combination of the MRTOF-MS and the $\alpha$-TOF detector, especially as mass and energy resolutions improve, will make a significant contribution to the investigation of the level structure of nuclides even for rare event cases such as the superheavy nuclides.

\section*{ACKNOWLEDGMENTS}
We express our gratitude to the RIKEN Nishina Center for Accelerator-based Science and the Center for Nuclear Science at University of Tokyo for their support of online measurements. Additionally, T. N. wishes to thank the RIKEN Junior Research Associate Program.
This work was financially supported by the Japan Society for the Promotion of Science KAKENHI (Grant No. 17H06090).

%------ reference from here ------- 

%\bibliographystyle{apsrev4-1}
%\bibliography{206207Ra}

\end{document}